\documentclass[aps,amsmath,prl,twocolumn,superscriptaddress,nofootinbib]{revtex4-2}
\usepackage{graphicx}
\usepackage{bm}
\usepackage{epsfig}
\usepackage{amsmath,latexsym}

\newcommand{\nn}{\nonumber}

\newcommand{\beq}{\begin{equation}}
\newcommand{\eeq}{\end{equation}}
\newcommand{\bea}{\begin{eqnarray}}
\newcommand{\eea}{\end{eqnarray}}

\begin{document}

\title{First Principle Predictions for   Cold Fermionic Gases Near Criticality via Critical Boson Dominance and Anomaly Matching}
\author{$^{1}$Shashin Pavaskar 
and $^{2}$Ira Z. Rothstein \vspace{0.7 cm} \\ $^{1}$ Dept. of Physics, Univ. of Illinois at Urbana-Champaign, Urbana, IL 61801, USA\\$^{2}$ Dept. of Physics, Carnegie Mellon University, Pittsburgh, Pennsylvania, 15213, USA}
%\date{} % Activate to display a given date or no date (if empty),
         % otherwise the current date is printed 

%%%%%%%%%%%%%%%%%%%%%%%%%%%%%%%%%%%%%%%%%%

\begin{abstract}
Recently the authors have developed an effective field theory formalism to systematically describe 
cold fermionic gases near the unitary limit.
The theory has enhanced predictive power due to the fact that interactions are dominated by the exchange of a gapped critical boson whose
couplings  and mass are fixed by matching  the dilatation anomaly between the UV and IR theories.
 We utilize this theory to give analytic   predictions for the compressibility and magnetic susceptibility for fermions near unitarity with attractive interactions above the critical temperature $T_c$,  with a well defined
theoretical error. The inputs to the predictions are: the scattering length $a$, the effective mass $m^\star$ and contact parameter $\tilde C(a)$.
We then compare  our predictions to numerical simulations and find excellent agreement within the window of scattering lengths where the EFT is valid ($10\geq \mid \!  k_F a \! \mid\geq 1$).  Experimental corroboration of this theory supports  critical point that can be describe by the inclusion of a scalar dilaton mode, whose action is fixed by symmetries.

%  in the  strongly interacting unitary phase  such systems may be described by actions which 
%may or may not contain an explicit field for the dilaton, the Goldstone boson associated with  the breaking of the non-relativistic conformal (Schrodinger) symmetry by the finite chemical potential.
%We  
%utilize  anomaly matching  to predict the quasi-particle width, away from unitarity, as a function of the contact parameter and the scattering length, . 
%We use this relation to  predict that the quasi-particle width is given by the expression $\Gamma(E,T)= \frac{8m}{9\pi {\tilde {\cal C}}^2 }\left(\sqrt{ \frac{m}{m^\star}}\frac{ a\mu_F^2}{4\hbar E_F^2} \right)^2 (E^2 +  (\pi kT) ^2)   ,$
%where $a$ is the the scattering length, $m_\star$ the effective mass  and  $\tilde {\cal C}$ is the dimensionless contact parameter .  This prediction is valid for $\left( \frac{E_F}{E}\right)^2  \gg a k_f\gg 1$ 
%On the other hand, should the dilaton action not be the appropriate
%description, then we predict that at unitarity ARPES experiments should yield a quadratic dispersion relation with $E =\frac{ p^2}{2m^\star}$ for the degrees of freedom which couple to the number density.

\end{abstract}

\maketitle
\section{Introduction}

Non-Fermi liquid  (NFL) behavior is  a  paradigm of exotic quantum phenomena  that, to date, is still lacking a theoretical underpinning.  It is usually assumed that deviations from canonical Fermi liquid behavior near quantum critical points are
due to relevant fluctuations of ``critical bosons" corresponding to fluctuations in an order parameter which for broken continuous symmetries will be Goldstone bosons. For systems with unbroken space-time symmetries, Goldstones are typically irrelevant in the infra-red since they couple derivatively, but in condensed matter systems, this no longer need be the case.
 Such non-derivatively coupled Goldstones will in general drive the system to strong coupling in the infra-red.
  The symmetry breaking condition sufficient for such behavior were determined in the non-relativistic case in \cite{vish} and later generalized to the relativistic case in \cite{RS2}.

Describing the behavior of a field theory at any critical point presents a strong coupling challenge with no obvious expansion parameter with which to control the calculation. The complexity of the critical  system is  amplified, for interacting fermions, due to  the presence of a lattice.   Thus to gain a better understanding of such systems it is prudent to consider the simpler case of a gas of fermions, not only for the sake of simplicity, but also to utilize the
experimental power  to precisely engineer such systems. The ability to fix  the scattering length of the microscopic interactions between neutral particles, allows one to tune cold fermionic systems to a critical point, leading to 
a non-relativistic conformal field theory manifesting the symmetry of the full Schrodinger group. 
This critical point controls the cross-over between the BEC and the BCS sides of the phase diagram.
It has been proven that at this  point the system can not be described as a canonical Fermi liquid \cite{RS}, that is, the critical system falls under the rubric of a NFL.
The basic reason for this behaviour
can be ascribed to the fact there  is no way to non-linearly realize the broken space-time symmetries of the system without the 
introduction of a critical boson whose fluctuations dominate at long distances due to the existence of the aforementioned non-derivative couplings.

While there is evidence that  thermodynamics properties of unitary fermionic gases behave similarly to fermi liquids, 
microscopic  properties, such as the spectral functions have recently been shown to deviate from Fermi liquid behaviour.
 In  particular, in \cite{Li}, in a box trap, the spectral function demonstrated the characteristics of the ``pseudo-gap'' above the critical temperature.

We can determine the quantum numbers and couplings of the critical boson by studying the symmetry breaking pattern.
If we consider the UV theory as the quantum mechanics of particles scattering at infinite scattering length then, as mentioned, 
the symmetries of the systems are enhanced to those of the Schrodinger group. When the system in placed in a vacuum with
finite chemical potential, the conformal symmetries (dilatations and special conformal transformations)  as well as Galilean boosts are  spontaneously broken resulting in a collection of Goldstone
bosons, though the actual number of modes need not be equal to the number of broken generators. If we deform away from the critical point, by making the scattering length large but finite, then the Goldstones will get
gapped and we can again describe the system as a Fermi liquid. As long the Goldstone boson mass are small compared to UV scales, we can still utilize the (approximate) symmetries to write down the form of the action. As was shown in \cite{RS,RS2} all of the symmetries can be realized with only the need for one (pseudo) Goldstone boson,  which we will call the ``dilaton".

To understand the impact of a dilaton on the dynamics we first consider the effective field theory of a canonical Fermi liquid 
which would arise if we are sufficiently far away from the critical point.
In this EFT \cite{Shankar,Polchinski} we expand all momenta around the Fermi momenta $\vec{k}_F$, such that $k/k_F \ll 1$.
 The relevant degrees of freedom are electron 
quasi-particles  interacting via a contact potential. Expanding around the Fermi surface the leading order action is given by
\begin{eqnarray}
S_{\psi} &=& \int d^{3}x  \hspace{0.05cm} dt \hspace{0.1cm}  (i\psi^{\dagger} \partial_{t}\psi + \psi^{\dagger}\vec{v}_{F}\cdot \vec{\partial}\psi +  g_{BCS}(\theta)\psi^{\dagger} \psi \psi^{\dagger} \psi \nn \\ &+& g_{FS}(\theta)\psi^{\dagger} \psi \psi^{\dagger} \psi),
\end{eqnarray}
where for simplicity we have suppressed the spin indices.
The Fermi velocity is defined as $v_F=  \hat n \cdot \partial \epsilon(k)/\partial \vec k $, where $\hat n$ is the normal to the
Fermi surface. The two leading order interactions have very specific kinematics.
In the BCS channel the scattering is purely back to back as opposed to the forward scattering ($FS$) interaction.
These are the only interactions that conserve  momentum and leave all states near the Fermi surface \cite{eft,Shankar,Polchinski}.
Alternatively, these are the only interactions consistent with the invariance under space-time symmetries \cite{RS}. The coupling functions $g_{BCS}$ and $g_{FS}$ only depend upon one scattering angle given the rotational invariance of the system since, to leading order in the expansion of $k/k_F$, we can set the magnitude of all the momenta in the
coupling function to $k_F$. 
It is common practice to decompose these couplings functions into partial waves.
This action realizes all of the broken space-time symmetries which are the three Galilean boosts.
As long as the the Landau
 relation between $g_{FS}(l=1)$
 and the effective mass $m_\star$ defined by $m_\star= k_F/v_F$ is satisfied, there is no need for a boost Goldstone
 boson \cite{RS2}.
 
 This action can not properly describe the critical point as it is not manifestly invariant under the
 spontaneously broken Schrodinger symmetry group.  However, by adding a dilaton to the action with the proper couplings, one can repristinate the full conformal symmetry albeit as a non-linear realization. The appropriate action is given by
\begin{eqnarray}
     S_{\psi}= \int dt  d^{3}x \:  {\psi^{\dagger}}( i\partial_{t}{\psi} - e^{\frac{2\phi}{\Lambda}}\epsilon(e^{-\frac{\phi}{\Lambda}}i\vec{\partial}){\psi}  ) +  \frac{f_{FS}}{2} ( \psi^\dagger  \psi)^2 
   ] , \nn \\
 \end{eqnarray}  where 
%spin indices will be suppressed and  
  $\Lambda$ is the conformal symmetry breaking scale which is set by the chemical potential.
 
 Expanding  around the Fermi surface to leading order, the dilaton field $\phi$ interaction   is given by, 
\begin{equation}
\label{action}
S = \int d^{3}x  \hspace{0.05cm} dt \hspace{0.1cm}   \frac{\phi}{\Lambda}\psi^{\dagger} \psi( 2 \epsilon(k_F)- \vec \partial_{p}\epsilon(k_F) \cdot \vec k_F)+...
\end{equation}
where we have dropped terms sub-leading in the power expansion, since momenta normal to the Fermi surface scale as $k/k_F$.
%Notice that the coupling vanishes in the free limit where $\epsilon=\vec p^2/(2m)$.
It is convenient to re-express this coupling in terms 
the Fermi velocity  $\vec v_F \equiv \vec \partial_{p}\epsilon(k_F)$ then
\begin{equation}
S_{\psi} = \int d^{3}x  \hspace{0.05cm} dt \hspace{0.1cm}   \frac{\phi}{\Lambda}\psi^{\dagger} \psi \left(  2 \epsilon(k_F)- v_F k_F \right)+...
\end{equation}
Notice that in the free limit, where the dispersion relation becomes quadratic $\epsilon=k^2/(2m)$, the dilaton decouples as it must.
For notational convenience we define  
\beq
\label{delta}
 \delta E\equiv \left(  2 \epsilon(k_F)- v_F k_F \right).
\eeq

 The action (\ref{action}) has the correct symmetries to describe the critical theory. 
 However, the system lacks a well defined notion of a quasi-particle as the fermion lifetime will be parametrically smaller than
 its inverse energy, i.e. the system will be a non-Fermi liquid.
 At present it is not known how to  maintain calculation control over such a system.
 
 Next we will deform away from this critical point by 
 allowing the microscopic theory to have a finite scattering length.
As long as the scattering length is such that $k_F a$ is not too large (to be quantified below) then this mild symmetry breaking will gap the dilaton 
 and  leads to a mass term for the dilaton \footnote{We are working in units 
 where the electron mass is one and $\hbar=1$, such that all units are measured in length.}
 \beq \delta L= -\frac{1}{2} \bar m_\phi ^2 v_D^4 \phi^2 . \eeq
 where $m_\phi$ and $v_D$ are the dilaton mass and velocity respectively.
 Now the upshot is that if the dilaton mass is sufficiently light, then the dilaton exchange will be enhanced relative
 to the contact term. To see this consider the dilaton exchange contibution to the electron quasi-particle scattering amplitude 
 \beq
 \label{amp}
 M \sim  \frac{1}{\Lambda^2(E^2-\vec p^2 v_D^2 -m_\phi ^2)}.
 \eeq
 where we have absorbed $v_D$ into the mass since it will cancel in our results below.
 As long as we are working at sufficiently low energy\footnote{Recall that the interaction is only marginal if the scattering happens in the forward direction.} (temperatures) then the interaction localizes
 \beq
 M \sim \frac{1}{(\Lambda  m_\phi)^2}
 \eeq
 and thus will be enhanced as we approach the critical point where the dilaton becomes massless.

Nonetheless, it would seem that we have gained no predictive power as we have simply traded one unknown coupling $f$ for
another, $ m_\phi$. However, as was pointed out in \cite{paper1}, anomaly matching allows us to fix 
$ m_\phi$ completely in terms of $k_f a, m^\star$ and the contact density ${\cal C}(a)$.
%This relation relies on the fact that the UV theory of interacting Fermions  in the trivial vacuum, is exactly solvable 
(See appendix for details)
 
 	\begin{eqnarray}
 	\label{money}
 	m_{\phi}^{2}\Lambda^{2}  &= - \frac{3}{4\pi a} \hspace{0.1cm} {\cal C}(a).
 \end{eqnarray}

If the dilaton mass is sufficiently small it will dominate the quasi-particles interactions, as other contributions  to the  interaction, arising from integrating out other modes, will be parametrically  suppressed by powers of $m_\phi/E_F$.

\section{Predictions}
%   In this section, we make a prediction for the decay width for fermions at large scattering lengths. In the EFT for a fermi liquid, the only relevant four-fermi couplings are forward scattering and BCS couplings. 
 The result in (\ref{money}) allows us to predict the value of the s-wave Landau parameter which is usually an unknown that is sensitive to  short distance strong coupling many-body physics. By working at energies below the scale of the dilaton mass (see \cite{paper1} for details), the dilaton exchange localizes and generates an effective s-wave
 Landau parameter
\beq
f_d\equiv \frac{4\pi a \delta E^2}{3 \tilde {\cal{C}}(a)k_F^4},
\eeq
such that the total local interaction coupling can be written as
\beq
\label{res}
 f_0= f_{FS}+f_d.
\eeq
 Note that we have replaced the contact density by the dimensionless contact parameter $\tilde {\cal{C}}(a)$ . The $f_d$  will dominate as long as the dilaton is sufficiently light.
 This will be quantified below.
 Since the mass of the dilaton scales as $(k_F a)^{-1}$,  dilaton domination places a lower bound on  $k_F a$, while an upper bound exists to make
 sure that the dilaton interaction localizes  (i.e. $E<m_\phi$ in (\ref{amp})).
 \beq
\left( \frac{E_F}{E}\right)^2 \tilde {\cal C}(a)> k_Fa > \left( \frac{E_F}{\delta E}\right)^2  \tilde {\cal{C}}(a)k_F f_{FS},
\eeq
where $k_F f_{FS} \sim 1$\footnote{This follows from the fact that $1/f_{FS}$ is the cut off of the EFT which suppresses all higher dimensional operators.}
Note that the BCS interaction, $g_{BCS}$, will not play a roll in our predictions below, so it can be ignored.

Before we make our prediction, we first nail down the numerical range of $k_F a$ for which we can trust our results. For $\tilde {\cal C}(a)$, we used the results in \cite{zwerger}. For $\delta E$ we use the relation
$v_F=\frac{k_f}{m^\star}$,  where the effective mass $(m^\star)$ at unitarity was measured in  \cite{Nir} to be $m^\star/m=1.13$.
Since we are working below this limit, we will take $m^\star/m \approx1.1$. 

Using (\ref{delta}) we have
\beq
\label{delta2}
\delta E=k_F^2(\frac{1}{m}-\frac{1}{m^\star}).
\eeq
Now since we are interested in thermodynamics quantities in this paper, using (\ref{delta2}) we can re-write the entire allowed range as
\beq
\left(\frac{ T_F}{T}\right)^2 \tilde{\cal{C}}(a)>k_F a  > \frac{\tilde{\cal{C}}(a)}{4(1-\frac{m}{m^\star})^2}.
\eeq
Since we are working in the unbroken phase we have $T>T_c$, and
numerical simulations  \cite{numerics,numerics2} indicate $T_c(k_Fa)$ drops off linearly for $-1\gtrsim (k_fa)^{-1}\gtrsim-.1$ peaking at unitarity $T_c/T_F\approx .15$.  We find that a self-consistent range of allowed values of $1/(k_F a)$ is roughly given by
\beq
1\gtrsim(k_F \mid a \mid )^{-1}\gtrsim .1
\eeq

%Comparing this to the self-energy for a normal fermi-liquid, we have 
%\begin{equation}
 %\begin{split}
    % \frac{\Sigma(\omega,\vec{k})}{\Sigma_{0}(\omega,\vec{k})} = \frac{f_{ren}^{2}}{f^{2}} &\sim 1 + 2\frac{\mu^{2}}{f*M}\\
%&\sim 1 + \frac{20\pi^{2} \mu^{2} }{3\zeta k_{F}^{4} }\left( \frac{I}{\Sigma_{0}}\right)^{\frac{1}{2}} a
% \end{split}
% \end{equation}

%One can see that the self energy of fermions at large scattering length differs from that of an ordinary fermi liquid by an additional term on the RHS. Since we can measure the self-energy of a fermi-liquid $\Sigma_{0}(\omega)$ in the laboratory and $I(\omega)$ is a known function of the energy, the relative change in the width as a function of the scattering length can be determined.  
%\begin{equation}
   %\frac{\delta\Sigma(\omega,\vec{k})}{\Sigma_{0}(\omega,\vec{k})} = \frac{20\pi^{2} \mu^{2} }{3\zeta k_{F}^{4} }\left( \frac{I}{\Sigma_{0}}\right)^{\frac{1}{2}} a
 %\end{equation}
 \begin{figure}
	\centering
	\includegraphics[angle=-90,width=1.\linewidth]{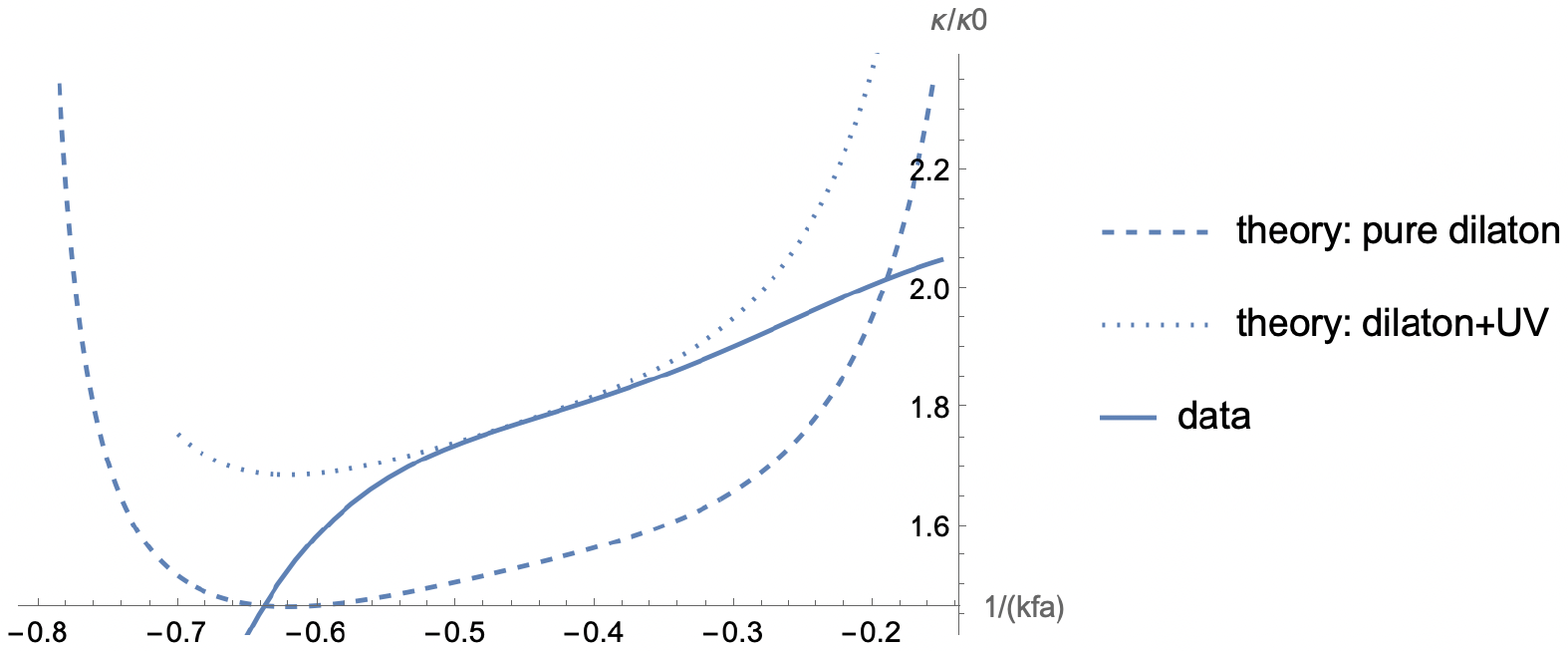}
	\caption{This plots compares the theory prediction for the compressibility in the range of the validity of the EFT, $10>\mid k_f a \mid >1$ to the numerical results in \cite{numerics}.  The dotted line shows the theory predictions with no free parameters, based solely on the values of  $k_f a,m^\star$ and the contact parameter. We see that it matches the numerical data for the slope very well and the normalization is off by approximately $\% 20$  in accordance with the error budget. If one fits the value of the the  short distance contribution $f$ at one value of $a$, we see that the prediction lies on top of the data for within the range where the EFT is valid. }
	\label{fig:Validity of the EFT}
\end{figure}

Given our prediction for the coupling we now use the fact that the EFT gives a non-perturbative prediction for
the compressibility \cite{Shankar} given by 
\beq
\frac{\kappa}{\kappa_0}=\frac{1-\frac{\pi^2}{12} \frac{T^2}{ T_F^2}}{1+\Big(1-\frac{\pi^2}{12} \frac{T^2}{ T_F^2}\Big) \frac{m^\star k_F}{m\pi^2}  f_{0}},
\eeq
where $\kappa_0$ is the compressibility of the free Fermi gas at zero temperature.
Using our result for the Landau parameter $(\ref{res})$, we can first consider only the contribution from $f_d$%we can consider the result first using the leading order
%prediction for $f$ which includes only $f_d$
, which makes a prediction for the normalization. 
The prediction using only
$f_d$ at $T_c$ is shown as the dashed line in Fig. 1. The scaling of the errors due to the UV contribution $f_{FS}$ is given
by 
\beq
f_{FS}/f_d \sim \frac{4\pi}{3} \frac{k_F a}{\tilde{\cal C}}\left(1-\frac{m}{m^\star}\right)^2 \sim .2
\eeq
where on dimensional grounds we took $f_{FS} \sim k_F^{-1}$ and  used  $\tilde{\cal C}\sim .1$. This ratio has some mild $a$ dependence, 
but importantly $f_{FS}$ does not. This error budget is consistent with figure one which shows that our prediction deviates from 
the data at the $\%20$ level. Given that $f_{FS}$ has no \footnote{It is possible that $f_{FS}$ has logarithmic $a$ dependence since
it is a marginal parameters that could be mildly sensitive to $a$, which would be a small correction to a small correction.}
 $k_Fa $ dependence, we can fit for it at one value of $k_F a$ and predict the rest of the plot, which is shown as the dotted line
 in Fig. 1. We see that our predictions lies on top of the numerical data in the heart of the region of validity of the effective theory. Additionally, the compressibility of the fermi gas for different temperatures has been plotted in Fig. 2, which also seem to agree well with the numerical predictions.
 
 Similarly one can predict the response to an external magnetic field and calculate the spin susceptibility of the interacting Fermi gas.
\beq
\frac{\chi}{\chi_0}=\frac{1-\frac{\pi^2}{12} \frac{T^2}{ T_F^2}}{1-\Big(1-\frac{\pi^2}{12} \frac{T^2}{ T_F^2}\Big) \frac{m^\star k_F}{m\pi^2}  f_{0}},
\eeq
where $\chi_0$ is the spin susceptibility of the non-interacting Fermi gas at zero temperature. This has been plotted in Fig. 3. for different temperatures.

\begin{figure}
\centering
\includegraphics[width=0.7\linewidth]{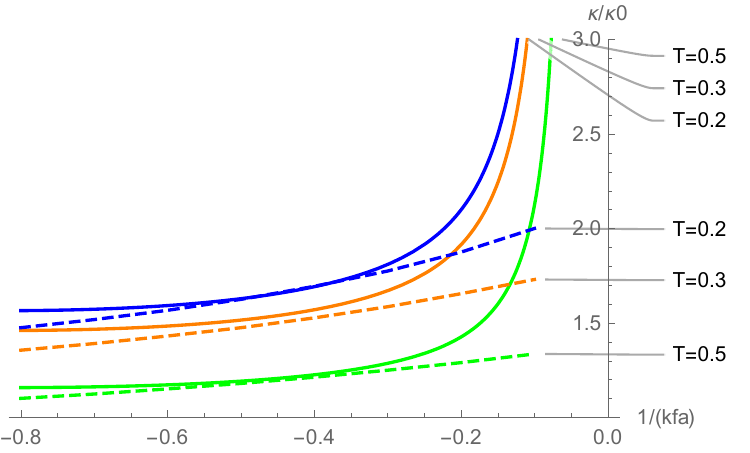}
\caption{This plot shows the compressibility of the Fermi gas as a function of $k_f a$ for various values of $T/T_f$. The solid lines depict the predictions of the EFT whereas the dotted lines show the numerical prediction from \cite{numerics}.  We have plotted the compressibility for three different values of $T/T_f$.  }
\end{figure}

\begin{figure}
	\centering
	\includegraphics[width=0.7\linewidth]{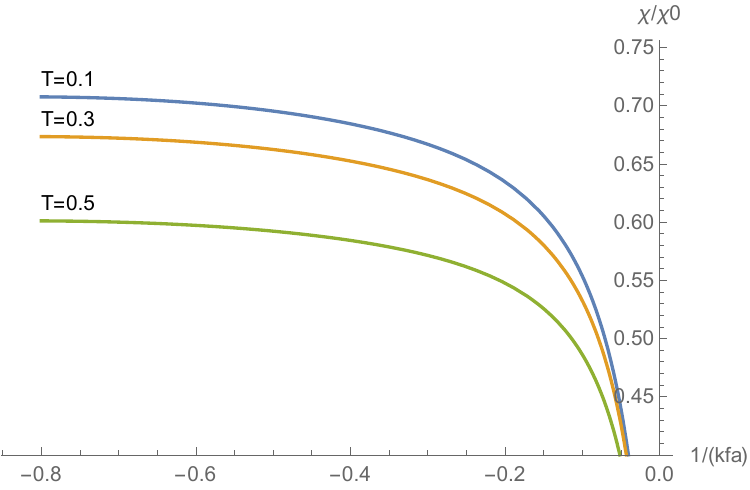}
	\caption{The above plot shows the susceptibility $\chi$ of the Fermi gas as a function of $k_f a$ for various values of $T/T_f$.  This is an independent prediction of our effective theory.}
\end{figure}

\section{Conclusions}

In this paper we have shown that if one can tune a system to be near a quantum critical point then the critical boson gets gapped but still dominates the low energy interactions and  allows us to predict the s-wave Landau parameter which dominates low energy physics.
This leads to a set of predictions which  are valid in a range of $10>k_f a > 1$ where the dilaton dominates 
the local interaction between quasi-particles. 

Our predictive power is predicated on our ability to fix  the mass of the dilaton in terms of the scattering length and contact parameter. This  is accomplished  by matching the current algebra in the effective theory to that of the full theory.  This in turn allows us to predict the compressibility and the magnetic susceptibility of the Fermi gas in the strongly interacting regime. 
% which in turn allows us to make a prediction for the quasi-particle lifetime including the normalization. The width is predicted to scale quadratically with the  ratio of scattering length to the contact parameter.
Also note that, since the dilaton is non-derivatively coupled, it will only generate the $l=0$ Landau parameter.
Thus we have the additional prediction that the s-wave Landau parameter will dominate all other channels.
These  predictions have a limited range of validity since the energy must be small enough that the dilaton exchange can still 
be treated as a local interaction. This limitation also implies our EFT breaks down when the scattering length, which is inversely proportional to
the dilaton mass,  becomes  too large.
Agreement with the data gives support to the idea that the unitary fermi gas at the critical point can be described
by coupling the electron field to a dilaton that leads to a NFL. To the best of our knowledge this is the first
evidence for a dilaton in nature.

\section{Acknowledgments}
We thank  the authors of \cite{numerics} for sharing their data with us.  We also benefitted from multiple discussion with Nir Navon. This work was partially supported by the US Department of Energy under grants   DE- FG02-04ER41338 and FG02-06ER41449.  
    
\section{Appendix A: Anomaly Matching}
	Here we present the details of the anomaly matching used to obtain the mass of the dilaton. The UV theory of interacting fermions in the trivial vacuum is exactly solvable. This allows one to calculate matrix elements in the UV theory and match them to those in the effective theory. The action for this theory is given by
\begin{equation}
	S = \int dt \int d^{3}x \hspace{0.1cm}i \chi^{\dagger} \partial_{t} \chi +  \frac{1}{2m}\chi^{\dagger} \nabla^{2} \chi - g(\mu) (\chi^{\dagger}\chi)^2
\end{equation}
where $\chi$ is two-spinor. The Van der Waals scale($\Lambda_{VDW}$) provides the upper cutoff in the theory that suppresses higher dimensional operators not shown.  At sufficiently low energies, the two-particle scattering process is dominated by the s-wave interaction, which is all that has been included here. The phase shift depends on the scattering length and the range ($R$) of the two-body potential ($R$ $\sim 1/\Lambda_{VDW} $). In the limit of $a>>R$ the renormalized coupling can be written in terms of the scattering length as
\cite{Braaten:2004rn}
\begin{equation}
	g(\mu) = \frac{4\pi}{-\frac{2}{\pi}\mu + \frac{1}{a}},
\end{equation}
where $\mu$ is the renormalization scale. This result is exact up to finite range corrections.

The divergence of the dilatation current is given by
\begin{equation}
	\partial_{\mu}s^{\mu} = (g(\mu) + \beta(g))(\chi^{\dagger}\chi)^2
\end{equation}
where $\beta= \mu \frac{d g}{d\mu}$, and the divergence vanishes at criticality. 
Our eventual goal is to match the vacuum matrix element of this operator equation with the corresponding
divergence operator equation in the effective theory which is given by
\begin{equation}
	\partial_{\mu}s^{\mu} =- m_{\phi}^{2}\Lambda \hspace{0.1cm}\phi.
	%+ \frac{2\phi^{2}}{\Lambda} +..)
\end{equation}
However, in doing so we would get relation between the vacuum expectation values of the four Fermi operator and
the dilaton which is not the object of interest. We want a relation that picks out only the combination $m_\phi \Lambda$ since that is the combination which shows up in the scattering amplitude (\ref{amp}).
To accomplish this, we note that we know that the dilaton, being a Goldstone boson, shifts by a constant under the action of the dilation generator which is given in the UV theory by 
\begin{eqnarray}
	D^{0}(0)  
	%\int d^{3}x \hspace{0.1 cm} s^{0}(\vec{x},0) 
	=  \int d^{3}x (\frac{3}{2}\chi^{\dagger}(\vec{x},0)\chi(\vec{x},0) + \chi^{\dagger}(\vec{x},0)\vec{x}\cdot \vec{\partial}\chi(\vec{x},0)) \nn \\
\end{eqnarray}
and  leads to
\begin{eqnarray} \label{algebra}
	[D^{0}(0),    \int d^{3}y(\partial_{\mu}s^{\mu}(\vec{y},0))] \!\!&=&\!\! 3\int d^{3}x \hspace{0.1cm} (g(\mu)+ \beta(g)) \hspace{0.1cm}(\chi^{\dagger}\chi)^2. \nn \\
\end{eqnarray}
While in the IR theory the dilaton generator is given by
\begin{equation}
	D^{0}(0) = \Lambda \int d^{3}x \hspace{0.1 cm}  \pi(\vec{x},0)
\end{equation}
where $\pi(x)$ is the conjugate momentum to $\phi$. 
The result analogous to (\ref{algebra})  gives
\begin{equation}
	\begin{split}
		\int_{x} \hspace{0.1cm} [D^{0}(0),\partial_{\mu}s^{\mu}(\vec{x},0)] %&=  m_{\phi}^{2}\Lambda^{2} \int d^{3}x' \hspace{0.1cm} \int d^{3}x \hspace{0.1cm} [ \pi(\vec{x}',0),\phi(\vec{x},0)] \\ 
		&=  \int d^{3}x \hspace{0.1cm} m_{\phi}^{2}\Lambda^{2}, %\equiv \int d^{3}x \hspace{0.1cm} M.\\ 
	\end{split}
\end{equation}
and we are now in position to predict the combination of parameters $m_{\phi}^{2}\Lambda^{2}$.

Equating vacuum matrix elements gives
\begin{eqnarray}
	\label{money1}
	m_{\phi}^{2}\Lambda^{2}  &= &
	-  \frac{3}{4\pi a }\langle g^{2}  \hspace{0.1cm}   \chi_{\uparrow}^{\dagger}\chi_{\uparrow}   \chi_{\downarrow}^{\dagger}\chi_{\downarrow} \rangle
	\equiv - \frac{3}{4\pi a} \hspace{0.1cm} {\cal C}(a)
\end{eqnarray}
$\cal C$ is the  contact density \cite{Tan}  which depends upon the scattering length
and has been extracted experimentally (see below).
Note that  
\beq (g(\mu)+ \beta(g)) (\chi^{\dagger}\chi)^2= \frac{g^2}{4\pi a} (\chi^{\dagger}\chi)^2\eeq
is a renormalization group invariant which will be relevant later.
The relation in (\ref{money1}) is reminiscent of the Gell-Mann-Oakes-Renner  \cite{GOR} relation between
the meson masses and the quark  chiral condensate
\beq
f_\pi^2 m_\pi^2=2 (m_u+m_d) \langle \bar \psi \psi \rangle.
\eeq
where $f_\pi$ is the analogue of $\Lambda$ and is the scale of the spontaneous chiral symmetry breaking.
Other uses of this type of reasoning are in the Higgs pion coupling \cite{VZ,CCGGM} as well as the coupling of pions to
quarkonia \cite{GR}, both of which utilized the breaking of relativistic conformal symmetry to make predictions

\end{document}